\documentclass[12pt]{article}

\usepackage{hyperref}
\usepackage{cite}
\usepackage{units}
\usepackage{epsfig}
\usepackage{pifont}
\usepackage{amsmath,amsfonts,amssymb,amsthm,mathrsfs}
\usepackage{graphicx,chngcntr,slashed}
\usepackage{cancel}
\usepackage{pstricks}
\usepackage{afterpage}
\usepackage{braket}
\usepackage{caption}
\usepackage{subcaption}
\usepackage{xcolor}
\usepackage{verbatim}

\def\mathswitch#1{\relax\ifmmode#1\else$#1$\fi}

\makeatletter
\def\section{\@startsection {section}{1}{\z@}{+3.0ex plus +1ex minus
  +.2ex}{2.3ex plus .2ex}{\large\bf\boldmath}}
\def\subsection{\@startsection{subsection}{2}{\z@}{+2.5ex plus +1ex
minus +.2ex}{1.5ex plus .2ex}{\normalsize\bf\boldmath}}
\def\subsubsection{\@startsection{subsubsection}{3}{\z@}{+3.25ex plus
 +1ex minus +.2ex}{1.5ex plus .2ex}{\normalsize\it}}

\graphicspath{{.}{plots/}}

\begin{document}
\thispagestyle{empty}

\def\thefootnote{\fnsymbol{footnote}}

\begin{flushright}
\end{flushright}

\vspace{1cm}

\begin{center}

{\Large {\bf Four-lepton
    $Z$ boson decay constraints on the SMEFT}}
\\[3.5em]
{\large
Radja~Boughezal$^1$, Chien-Yi~Chen$^2$, Frank~Petriello$^{1,2}$ and Daniel~Wiegand$^{1,2}$
}

\vspace*{1cm}

{\sl
$^1$ HEP Division, Argonne National Laboratory, Argonne, Illinois 60439, USA \\[1ex]
$^2$ Department of Physics \& Astronomy, Northwestern University,\\ Evanston, Illinois 60208, USA
}

\end{center}

\vspace*{2.5cm}

\begin{abstract}

We discuss how four-lepton decays of the $Z$ boson probe currently unconstrained flat directions in the parameter space of the Standard Model Effective Field Theory (SMEFT). We derive the constraints from these decays on four-lepton operators in the SMEFT and show how the LHC data for this process complements probes from neutrino-trident production. Future differential measurements with high-luminosity data can strongly constrain four-lepton operators and remove all flat directions in the four-muon sector of the SMEFT. We comment briefly on the possibility of using rare $Z$-decays to $\tau$-leptons to probe  untested directions in the SMEFT parameter space.
  
\end{abstract}

\setcounter{page}{0}
\setcounter{footnote}{0}

\newpage


\section{Introduction}

The search for physics beyond the Standard Model (SM) at the Large
Hadron Collider (LHC) and other experiments has so far yielded no new
particles.  This lack of evidence for new electroweak-scale physics
suggests that there is a mass gap between the SM and the next energy
scale at which new particles appear.  Although the search for new
particles will continue in the future at the high-luminosity LHC, it
is becoming increasingly important to search for potentially small and subtle
indirect signatures of new physics, and to understand the constraints imposed by current data on high-scale new physics.  A systematic framework for
characterizing deviations from the SM in the presence of no new
electroweak-scale particle is the SM effective field theory (SMEFT).
The SMEFT is constructed by allowing higher-dimensional
operators containing only SM fields that respect the SM gauge
symmetries.  These operators are suppressed by an energy scale $\Lambda$ at
which the effective theory breaks down and new fields must be added to
the Lagrangian.  The leading lepton-number conserving dimension-6 operators
characterizing deviations from the SM have been
classified~\cite{Buchmuller:1985jz, Arzt:1994gp, Grzadkowski:2010es}.

Significant effort has been devoted to performing global
analyses of the available data within the SMEFT
framework with varying assumptions~\cite{Han:2004az,Pomarol:2013zra,Chen:2013kfa,Ellis:2014dva,Wells:2014pga,Falkowski:2014tna,deBlas:2016ojx,Cirigliano:2016nyn,Hartmann:2016pil,Falkowski:2017pss,Biekotter:2018rhp,Grojean:2018dqj,Hartland:2019bjb,Brivio:2019ius,Aoude:2020dwv}.
Since the general dimension-6 SMEFT Lagrangian contains 2499 parameters for three
generations assuming baryon-number conservation, quite often additional
flavor symmetries such as Minimal Flavor Violation (MFV) are assumed
in order to reduce the number of Wilson coefficients. Assuming MFV implies that the flavor structure of the SMEFT Wilson coefficients are carried by combinations of Yukawa matrices.  This leads to several familiar intuitions~\cite{Alonso:2013hga,Cullen:2020zof}: for example, that the coefficients of scalar and dipole operators are suppressed by small fermion masses for the lighter generations, and that vector four-fermion interactions are generation-independent.  Flavor assumptions such as MFV lead to several  advantages. In general fits without such flavor assumptions, flat directions exist since current
experimental constraints cannot access all possible Wilson coefficients. MFV also effectively suppresses strongly-constrained flavor-violating effects. 

Despite these advantages it remains important to extend fits within the SMEFT framework beyond the MFV assumption.  Going beyond MFV allows global fits to encompass a broader range of ultraviolet completions.  For example, models which attempt to explain discrepancies in rare $B$-meson decays have a structure that violates lepton flavor universality~\cite{Altmannshofer:2017fio}. Allowing for flavor structure in the SMEFT requires addressing and removing the flat directions between Wilson coefficients that appear. The removal of flat directions in fits to SMEFT Wilson coefficients require the use of additional processes and experiments~\cite{Boughezal:2020uwq}. In this work we point out that the rare $Z$ boson decays to four-leptons offer the potential to
probe combinations of four-fermion Wilson coefficients not accessible
in other measurements.  In particular, only a single combination of
four-muon Wilson coefficients is currently constrained in global fits
by the neutrino-trident production process $\gamma^{*} \nu_{\mu} \to \nu_{\mu}\mu^+\mu^-$~\cite{Altmannshofer:2014pba,Falkowski:2017pss}.  Four-muon
$Z$ boson decays at the LHC probe orthogonal combinations of these
Wilson coefficients, allowing for a complete determination of the four-muon operators in the SMEFT.  The potential of four-lepton decay modes to constrain physics beyond the SM has been investigated previously, particularly in the context of $Z^{\prime}$ models~\cite{Altmannshofer:2014pba,Rainbolt:2018axw}.
We study the constraints imposed by current LHC
data, as well as potential future constraints at a high-luminosity
LHC.  Current measurements of this mode consider only the total rate. We point out that differential measurements can completely determine all four-muon Wilson coefficients, which motivates their study with future high-luminosity data.  Although we focus here on the four-muon mode 
as experimental searches for $Z\to 4\mu$ exist, other channels such as 
$Z\to2\tau 2\mu$ and $Z\to 4\tau$ may provide probes of completely 
untested parameters in the SMEFT.  We comment briefly on this possibility in our conclusions.

Our paper is organized as follows.  We review aspects of the SMEFT needed for our analysis in Section~\ref{sec:smeft}. In Section~\ref{sec:inc} we study the constraints imposed by inclusive LHC measurements of the $Z \to 4 \mu$ decay rate on the SMEFT four-muon operators. We also discuss their complementarity with constraints from neutrino-trident production. We discuss what can be learned from future differential LHC measurements in Section~\ref{sec:diff}. Finally, we conclude in Section~\ref{sec:conc}.

\section{Review of the SMEFT} \label{sec:smeft}

We review in this section aspects of the SMEFT relevant for
our analysis of four-muon decays of the $Z$ boson.  The SMEFT is an extension of the SM Lagrangian to include terms
suppressed by an energy scale $\Lambda$ at which the ultraviolet completion
becomes important and new particles beyond the SM appear.  Truncating the expansion in $1/\Lambda$ at dimension-6, and
ignoring operators of odd-dimension which violate lepton number, we
have
\begin{equation}
{\cal L} = {\cal L}_{SM}+ \frac{1}{\Lambda^2}\sum_i C_{i} {\cal
  O}_{i} + \ldots,
\end{equation}
where the ellipsis denotes operators of higher dimensions.  The Wilson
coefficients $C_i$ defined above are dimensionless.  When
computing the $Z$ boson decay width we consider only the leading
interference of the SM amplitude with the dimension-6 contribution.
This is consistent with our truncation of the SMEFT expansion above,
since the dimension-6 squared contributions are formally the
same order in the $1/\Lambda$ expansion as the dimension-8 terms which we neglect. The Wilson coefficients are renormalization-scheme dependent quantities.  In an $\overline{\text{MS}}$ scheme they become scale-dependent and run with energy.  As we perform only a leading-order analysis in this manuscript we neglect this running.

Corrections to the $Z \to 4 l$ decay widths come from two
sources: shifts of the $Z \bar{l} l$ and $\gamma \bar{l}l$ vertices
that scale as $v^2/\Lambda^2$ where $v$ is the Higgs vev, and
four-fermion operators which scale as $E^2/\Lambda^2$ where $E$ is the
characteristic energy scale of the process. Note that the $\gamma \bar{l}l$ vertex is shifted from the SM expression in the $(G_{\mu},M_W,M_Z)$ input parameter scheme~\cite{Brivio:2017btx} adopted here since the electromagnetic coupling is shifted. We summarize in
Table~\ref{tab:ffops} the dimension-6 operators that shift the decay
width at leading-order in its perturbative expansion. 
\begin{table}[h!]
\centering
\begin{tabular}{|c|c||c|c|}
\hline
${\cal O}_{\substack{ll \\ prst}}$ & $(\bar{l}_p\gamma^{\mu} P_L l_r) (\bar{l}_s\gamma_{\mu} P_L l_t)$ 
& ${\cal O}_{\substack{le\\prst}}$ & $(\bar{l}_p\gamma^{\mu}P_L l_r) (\bar{l}_s\gamma_{\mu} P_R l_t)$ \\
${\cal O}_{\substack{ee\\prst}}$ & $(\bar{l}_p\gamma^{\mu} P_Rl_r)(\bar{l}_s\gamma_{\mu} P_Rl_t)$ 
&${\cal O}_{\phi WB}$ & $\phi^\dagger \tau^I \phi W_{\mu\nu}^I B^{\mu\nu}$  \\
${\cal O}_{\phi D}$&$(\phi^\dagger D^\mu \phi)^\ast(\phi^\dagger D_\mu \phi)$ 
&${\cal O}_{\substack{\phi e\\rs}}$ &$i(\phi^\dagger \overleftrightarrow{D}_\mu \phi)(\bar{l}_r \gamma^\mu P_R l_s)$ \\
${\cal O}_{\substack{\phi l\\rs}}^{(1)}$ &$i(\phi^\dagger \overleftrightarrow{D}_\mu \phi)(\bar{l}_r \gamma^\mu P_L l_s)$
&${\cal O}_{\substack{\phi l\\rs}}^{(3)}$ &$i(\phi^\dagger \overleftrightarrow{D^I}_\mu \phi)(\bar{l}_r\tau^I \gamma^\mu P_L l_s)$ \\
  \hline
\end{tabular}
\caption{Dimension-6 operators contributing to the decay $Z\to 4l$. The last five operators only lead to overall shifts of the SM $Z\overline{l}l$ and $\gamma\overline{l}l$ vertices.\label{tab:ffops}}
\end{table}
Here, $l$ denotes a Dirac lepton, $\phi$ the Higgs boson, and $W$ and $B$ the field-strength tensors of the SU(3)$\times$U(1) gauge bosons.  $p,r,s,t$ denote generation indices. We have introduced explicit projection
operators $P_{L,R}$ to denote the projections onto left-handed
doublets and right-handed singlets. All operators containing the Higgs field $\phi$ only shift the 
$Z\bar{l}l$ and $\gamma\bar{l}l$ vertices and can be combined into shift constants $\delta g_{Z}^L$, $\delta g_{Z}^R$, $\delta g_{\gamma}^L$ and $\delta g_{\gamma}^R$ respectively. We note that $\delta g_{\gamma}^R=\delta
g_{\gamma}^L = \delta g_{\gamma}$.  Explicit expressions for these shifts are given in Appendix~\ref{app:A}.

There are potentially additional dimension-6 contributions from operators modifying the total $Z$ boson decay width that appear in the denominator of the branching ratio. To study these effects we express each $Z$ boson partial width in terms of its dimension-4 and dimension-6 contribution:
\begin{equation}
    \Gamma_i = \Gamma_i^{(4)}+\Gamma_i^{(6)},
\end{equation}
where the dimension-6 contribution is assumed to be small.  The branching ratio for $Z \to 4\mu$ can then be written as
\begin{equation}
    \text{BR}(Z\to 4\mu) = \frac{\Gamma(Z\to 4\mu)}{\sum_i \Gamma_i}\approx \frac{\Gamma^{(4)}(Z\to 4\mu)}{\sum_i \Gamma_i^{(4)}}\left[ 1+\frac{\Gamma^{(6)}(Z\to 4\mu)}{\Gamma^{(4)}(Z\to 4\mu)}-\frac{\sum_i \Gamma_i^{(6)}}{\sum_i \Gamma_i^{(4)}}\right]
\end{equation}
where the sum over $i$ includes all $Z$ boson decay modes and we have expanded to linear order in the dimension-6 corrections.  The second term in the square bracket above comes from the dimension-6 corrections to the total decay width. Since $\sum_i \Gamma_i^{(4)} \gg \Gamma^{(4)}(Z\to 4\mu)$, the only significant corrections from this last term come from the large $Z$ partial decay widths.  We assume that the dominant corrections come from the $Z \to \bar{f}f$ decay widths.  The corrections to these widths come from shifts in the left and right-handed couplings of the $Z$ boson to the different fermions, and are analogous to the operators leading to the shifts of the leptonic vertices in Table~\ref{tab:ffops} above. We will study the effects from shifts to the $Z\to \bar{f}f$ decays and absorb them into global shift factors $\delta g^{L,R}_{Zf}$ for each fermion species.

Finally, we note that dipole operators that can potentially contribute vanish for massless fermions upon truncation of the EFT expansion to dimension-6, which we assume here. The explicit expressions for all decay widths in terms of the SMEFT coefficients can be found in Appendix~\ref{app:A}.


\section{Constraining the four-lepton operators} \label{sec:inc}

The coefficients parameterizing the SMEFT contributions to the decay $Z \to 4\mu$ depend on the matrix elements describing the process and the imposed experimental cuts.  We evaluate them numerically at leading-order using the {\tt Madgraph} package {\tt SMEFTsim}~\cite{Brivio:2017btx}.  The UV scale $\Lambda$ is set to the Higgs vacuum expectation value $v=246$ GeV for
comparison with previous results in the literature~\cite{Falkowski:2017pss}. Explicit expressions for the four-lepton decay widths are given in Appendix~\ref{app:B}. We express the deviation for the four-muon decay mode in terms of the normalized branching ratio 
\begin{align}
\frac{\text{BR}(Z \to 4 \mu)}{\text{BR}_{SM}} = 1&+a_{ll}
C_{\substack{ll \\ 2222}}+a_{le}  C_{\substack{le \\ 2222}}+a_{ee}
C_{\substack{ee\\2222}} +a_{Zl}^L \delta g_{Zl}^L +a_{Zl}^R \delta g_{Zl}^R
+a_{\gamma \mu} \delta g_{\gamma \mu} \nonumber\\
&+a_{Z\nu}^L \delta g_{Z\nu}^L+a_{Zu}^L \delta g_{Zu}^L+a_{Zu}^R \delta g_{Zu}^R+a_{d}^L \delta g_{Zd}^L+a_{d}^R \delta g_{Zd}^R ,
\label{eq:BRdef}
\end{align}  
where we assume lepton and quark flavor universality for the vertex shift operators for clarity of this argument. We note from the expression above that this decay is directly sensitive to the four-muon couplings in the SMEFT, making it of interest for accessing these
operators.  

We discuss next how well measurements of the inclusive $Z \to 4\mu$ decay
width can constrain the four-muon SMEFT operators defined in the
previous section.  We first show that after accounting for the strong constraints on the vertex shifts from $Z\to 2f$ decays at LEP and other experiments,
measurements of the $Z \to 4 l$ decay are primarily 
sensitive to leptonic four-fermion operators, which are not as strongly bounded yet.  The relevant comparison to establish 
whether $Z$ vertex shifts or four-fermion terms dominate the
SMEFT correction is the size of $a_i C_i$ for each term defined in
Eq.~(\ref{eq:BRdef}).  We evaluate
the branching ratio imposing $80\,\text{GeV}<m_{4l}<100\,\text{GeV}$
and $m_{ll}>4$ GeV for all fermion
pairs, consistent with experimental analyses, and obtain the 
results for the $a_i$ shown in Table~\ref{tab:ffopscoeff}.
\begin{table}[h!]
\centering
\begin{tabular}{|c|c|c|c|c|c|c|c|c|c|c|}
\hline
$a_{ll}$ & $a_{ee}$ & $a_{le}$ & $a_{Zl}^L$ & $a_{Zl}^R$  & $a_{\gamma\mu}$&$a^L_{Z\nu}$ &$a_{Zu}^L$&$a_{Zu}^R$&$a_{Zd}^L$&$a_{Zd}^R$ \\ 
\hline
0.025 & 0.016 & 0.009 & 4.2 & -3.3&4.0 &-0.40&-0.39&-0.071&-0.87&-0.027 \\  
  \hline
\end{tabular}
\caption{Results for the $a_i$ coefficients given the cuts $80\,\text{GeV}<m_{4l}<100\,\text{GeV}$
and $m_{ll}>4$ GeV. For comparison with the available $Z$-pole bounds we assume flavor universality for the vertex-shift operators.\label{tab:ffopscoeff}}
\end{table}
The lepton vertex-shift $a_{Zl}^{L,R}$ factors are in general two orders of magnitude larger
than the four-muon $a$ coefficients for the relevant experimental cuts.  However, the 
$\delta  g_{Zl}^{L,R}$ are constrained to be $2 \times 10^{-4}$ or
smaller from LEP data~\cite{ALEPH:2005ab}.  
The hadronic vertex shifts that enter the branching fraction through the total width are similarly constrained through the available LEP data, though the bounds are generally weaker than the ones on the leptonic coefficients by a factor $3$ to $5$. However, the shift factors $a_{Zu,d}^{L,R}$ are small, and these corrections are numerically negligible. The relevant bounds on the Wilson coefficients that enter the vertex shifts from the literature are adapted from~\cite{Dawson:2019clf} and are summarized in Table~\ref{tab:bounds}. The single combination of
four-muon couplings probed by neutrino trident production is only
constrained to the $2 \times 10^{-1}$ level~\cite{Falkowski:2017pss}, and the two orthogonal combinations are not bounded at all. Therefore the $a_i C_i$ factors for the four-muon couplings are
allowed to be at least an order of magnitude larger than those of the
vertex shifts.  In what follows we assume the vertex shifts are
strongly constrained by other measurements and neglect them,
consistent with the above observation.
\begin{table}[ht]
\begin{center}
 \begin{tabular}{|c| c|| c|c|} 
 \hline
 $|C_{\phi D}|$ & $<0.0012$&$|C_{\phi l}^{(1)}|$&$<0.0006$   \\ 
 $|C_{\phi WB}|$ & $<0.0017$  &$|C_{\phi l}^{(3)}|$&$<0.0029$ \\
 $|C_{ll}|$ & $<0.0006$& $|C_{\phi e}|$&$<0.0003$  \\
 $|C_{\phi q}^{(1)}|$ & $<0.0023$& $|C_{\phi q}^{(3)}|$&$<0.0005$  \\
 $|C_{\phi u}|$ & $<0.0073$& $|C_{\phi d}|$&$<0.014$  \\
 \hline
 \end{tabular}
 \caption{$68\%$ confidence-level (C.L.) bounds for a single Wilson coefficient. The UV scale is set to $\Lambda = 246\,\textrm{GeV}$.  The bounds are derived assuming flavor universality. These numbers are adapted from Ref.~\cite{Dawson:2019clf}. \label{tab:bounds}}
 \end{center}
\end{table}
\subsection{Single Wilson-coefficient constraints from inclusive LHC measurements}
Both ATLAS and CMS have performed measurements of the $Z \to 4l$
branching ratios~\cite{CMS:2012bw,Khachatryan:2016txa,Sirunyan:2017zjc,Aad:2014wra}.  These experiments are summarized in
Ref.~\cite{Rainbolt:2018axw}, where a combination of existing
results is also given.  The combined measurement of the
$Z \to 4l$  branching ratio is:
\begin{equation}
\text{BR}(Z \to 4l) = (4.58 \pm 0.26) \times 10^{-6}.
\end{equation}  
The measurements are scaled via a Monte-Carlo simulation to the
following common phase-space region:
\begin{equation}
80\, \text{GeV}<m_{4l}<100\, \text{GeV},\;\;\; m_{l^+l^-}>4\, \text{GeV}
\end{equation}  
where $m_{l^+l^-}$ refers to the invariant mass of any combination of
oppositely-charged leptons.  As our interest is in the four-muon mode,
we convert this combination to a result for $\text{BR}(Z \to
4\mu)$ as follows. For each of the experimental measurements in Table~1 of Ref.~\cite{Rainbolt:2018axw} we scale the central value by the leading-order ratio
$\Gamma(Z \to 4\mu)/\Gamma(Z \to 4l)$ computed using {\tt Madgraph},
and the statistical uncertainty by $\sqrt{\Gamma(Z \to 4l)/\Gamma(Z
  \to 4\mu)}$.  Combining the results yields
\begin{equation}
\text{BR}(Z \to 4\mu) = (1.21 \pm 0.41) \times 10^{-6}.
\end{equation}  
We have checked that including the correlation coefficients listed in
Table~2 of  Ref.~\cite{Rainbolt:2018axw} does not
significantly change this result.  We estimate the expected
constraints from an inclusive measurement at a high-luminosity LHC by
assuming that both statistical and systematic errors scale as
$1/\sqrt{{\cal L}}$, where ${\cal L}$ is the integrated luminosity.
We show the current constraints and those from a HL-LHC assuming 3000 fb$^{-1}$ below in
Table~\ref{tab:fit} for each four-muon Wilson coefficient turned on
separately.  The current constraints on the Wilson coefficients from the inclusive branching ratio measurement are quite weak.  They become much stronger with the full HL-LHC data set.   
\begin{table}[ht]
\centering
\begin{tabular}{|c||c|c|}
\hline
 & Current ($Z\to 4\mu$) &  HL-LHC ($Z\to 4\mu$) \\
\hline
$|C_{\substack{ll \\ 2222}}|$      & $<10.5 $     &$< 1.0 $     \\

$|C_{\substack{ee \\ 2222}}|
$       & $<16.9 $    &$< 1.6 $ \\

$|C_{\substack{le \\ 2222}}|
$    & $<28.8 $  &$< 2.7 $ \\
\hline
\end{tabular}
\caption{Single parameter constraints on the Wilson coefficients of the four-$\mu$ operators at 68\% CL for both the current LHC data and a projection
based on the HL-LHC with a luminosity of 3 ab$^{-1}$. For the
projection of the uncertainties at the HL-LHC, we assume that all  uncertainties scale as $\frac{1}{\sqrt{N}}$.
\label{tab:fit}}
\end{table}
We note that the choice of the phase-space constraint $m_{l^+l^-}$ was
not optimized for SMEFT studies.  However, the effect of increasing
this cut does not lead to stronger constraints with the current LHC
data.  We estimate this by using {\tt Madgraph} to compute the change
in branching ratio and consequently statistical error that occurs by
increasing the cut on $m_{l^+l^-}$. 
Although increasing the cut increases the size of the SMEFT-induced
deviation since it grows with energy, the corresponding increase in
the statistical error overwhelms this growth and leads to weaker bounds.

\subsection{Complementarity with neutrino trident production}
\label{sec:ntrident}

Another constraint on four-muon operators comes from the
neutrino-trident production process $\nu_{\mu} \gamma^{*} \to
\nu_{\mu} \mu^+\mu^-$ which occurs in the Coulomb field of a heavy nucleus.  Formulae for the deviation of this process
from SM predictions within the SMEFT framework are given in Ref.~\cite{Falkowski:2017pss}.  We reproduce this deviation below:
\begin{align}
    \frac{\sigma_\textrm{trident}}{\sigma_\textrm{trident}^\textrm{SM}} = 1 + \frac{2}{(1+4s_W^2+8s_W^4)}\frac{v^2}{\Lambda^2}&\left\{(C_{\substack{ll\\1221}} - C_{\substack{ll\\2222}}) (1 + 2 s_W^2) - 2 s_W^2 C_{\substack{le\\2222}} + 
 2 (\delta g^L_{W \mu}\right.\nonumber\\
 &+ \delta g^L_{Z\mu} - \delta g^L_{Z\nu_\mu} + 2s_W^2 \delta g^L_{W\mu} + 2 \delta g^L_{Z\mu} s_W^2 + 
    2s_W^2 \delta g^R_{Z\mu} \nonumber\\
    &\left.+ 8 s_W^4 \delta g^L_{Z\nu_\mu} -  (1 + 2 s_W^2)\delta g^L_{We})\right\}
\label{eq:ntrident}    
\end{align}
where $\delta g^L_{W\mu}$ is the shift to the $W\mu\nu_\mu$ vertex.  Its explicit expression in terms of standard SMEFT operators is given in Appendix~\ref{app:A}. We see that the deviation depends on the following combination of four-muon
Wilson coefficients:
\begin{equation}
\hat{C}_{\overset{ll}{2222}} = C_{\substack{ll\\2222}}+\frac{2
  s_W^2}{1+2s_W^2}  C_{\substack{le\\2222}}.
\end{equation}
From
this we see that this measurement is proportional to only a single
combination of $C_{\substack{ll\\2222}}$ and $C_{\substack{le\\2222}}$,
and is insensitive to $C_{\substack{ee\\2222}}$.  The $Z \to 4\mu$
decay is sensitive to all three operators in a different combination
than neutrino-trident production.  Once differential measurements are
made with higher luminosities, all three four-muon Wilson coefficients
can be separately determined from a combination of neutrino-trident production and LHC data.

\begin{figure}[htbp]
\centering
\includegraphics[width=.6\textwidth]{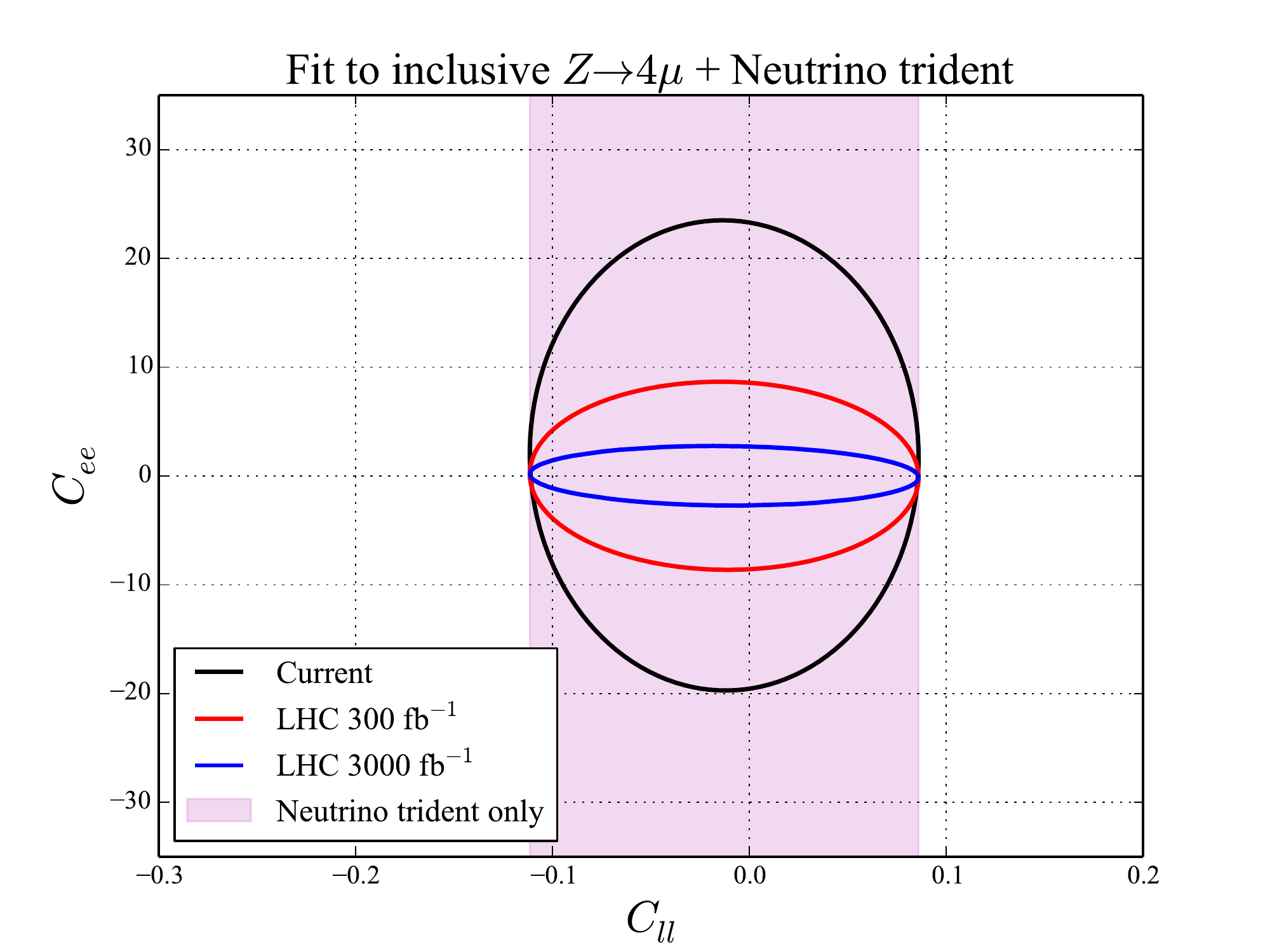}
\caption{68\% C.L. bounds on the combination of inclusive LHC $Z \to 4 \mu$ data and neutrino trident production assuming non-zero $C_{\overset{ll}{2222}}$ and $C_{\overset{ee}{2222}}$}
\label{fig:inccllcee}
\end{figure}

\begin{figure}[htbp]
\centering
\includegraphics[width=.6\textwidth]{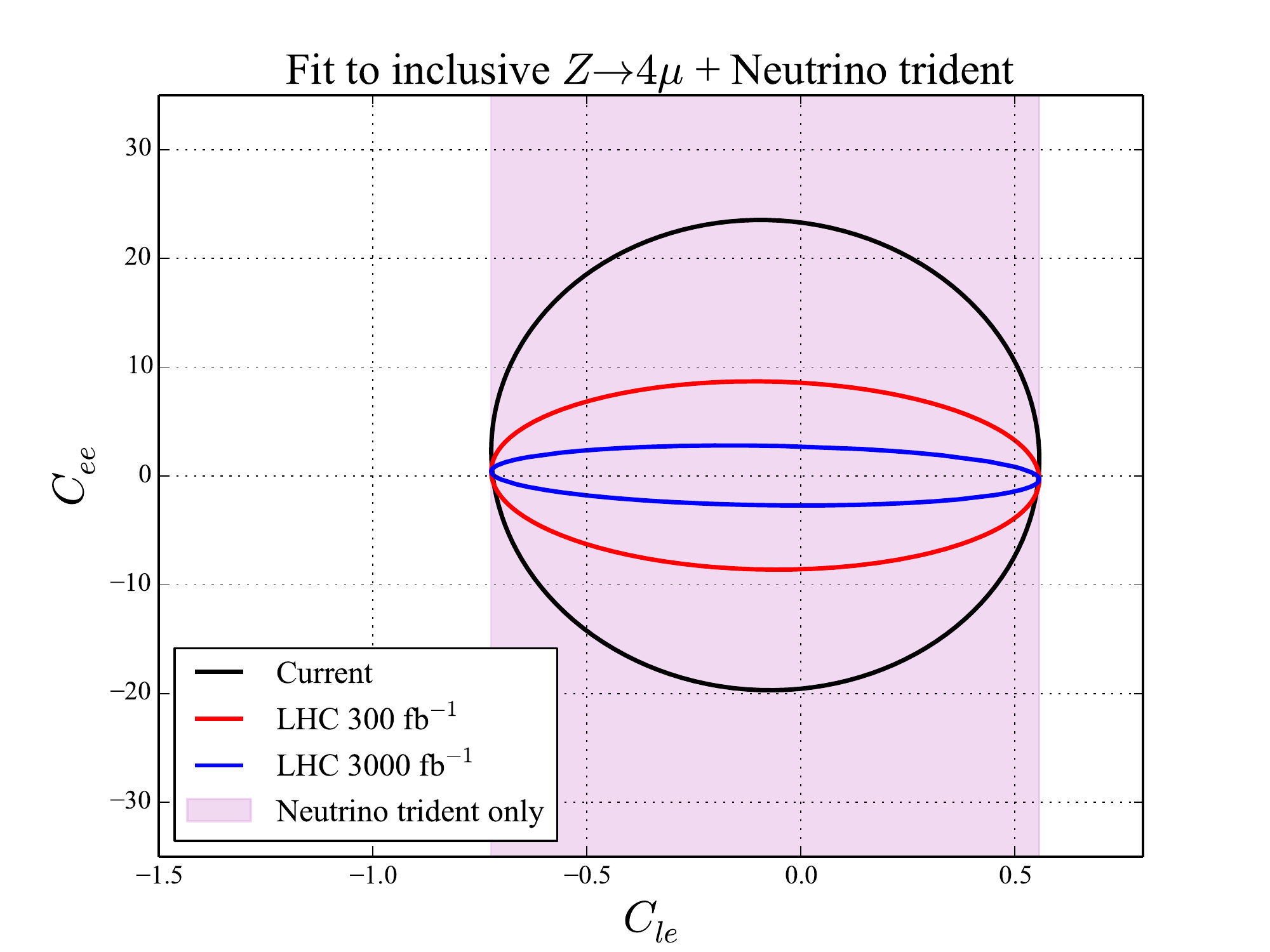}
\caption{68\% C.L. bounds on the combination of inclusive LHC $Z \to 4 \mu$ data
  and neutrino trident production assuming non-zero $C_{\overset{le}{2222}}$ and $C_{\overset{ee}{2222}}$ }
\label{fig:incclecee}
\end{figure}

To demonstrate what can be learned from a combination of neutrino-trident production and inclusive $Z \to 4\mu$ measurements at the LHC we perform fits to the inclusive LHC measurement
and neutrino-trident production data from the experiments CCFR~\cite{Mishra:1991bv} and CHARM-II~\cite{Geiregat:1990gz}. We consider two
different choices of Wilson coefficients:
\begin{enumerate}

\item $C_{\substack{ll\\2222}}$ and $C_{\substack{ee\\2222}}$
  non-zero;

  \item $C_{\substack{le\\2222}}$ and $C_{\substack{ee\\2222}}$
  non-zero.

\end{enumerate}  
The results of these fits are shown in Figs.~\ref{fig:inccllcee}
and~\ref{fig:incclecee}.  The solid bands refer to the constraints
from neutrino trident production.  We see that this data is not sensitive to $C_{\substack{ee\\2222}}$.
The ellipses refer to current LHC constraints, and projections for 300
fb$^{-1}$ and 3000 fb$^{-1}$ of integrated luminosity.  Including the LHC data removes the flat direction that occurs due to the insensitivity of neutrino-trident production to $C_{\substack{ee\\2222}}$.  We note that
the constraints from neutrino trident production on $C_{\substack{ll\\2222}}$ and $C_{\substack{le\\2222}}$ are stronger than the current LHC bounds, with these coefficients constrained to
be less than unity while current LHC data only requires
$C_{\substack{ee\\2222}}\lesssim 20$.  The power of the LHC measurement
increases with higher luminosities.  With 3000 fb$^{-1}$ the constraints
on $C_{\substack{ee\\2222}}$ approach the level of the neutrino-trident production bounds on $C_{\substack{ll\\2222}}$ and $C_{\substack{le\\2222}}$. 

\section{Differential measurements with future LHC data} \label{sec:diff}

\begin{figure}[htbp]
\centering
\includegraphics[width=.75\textwidth]{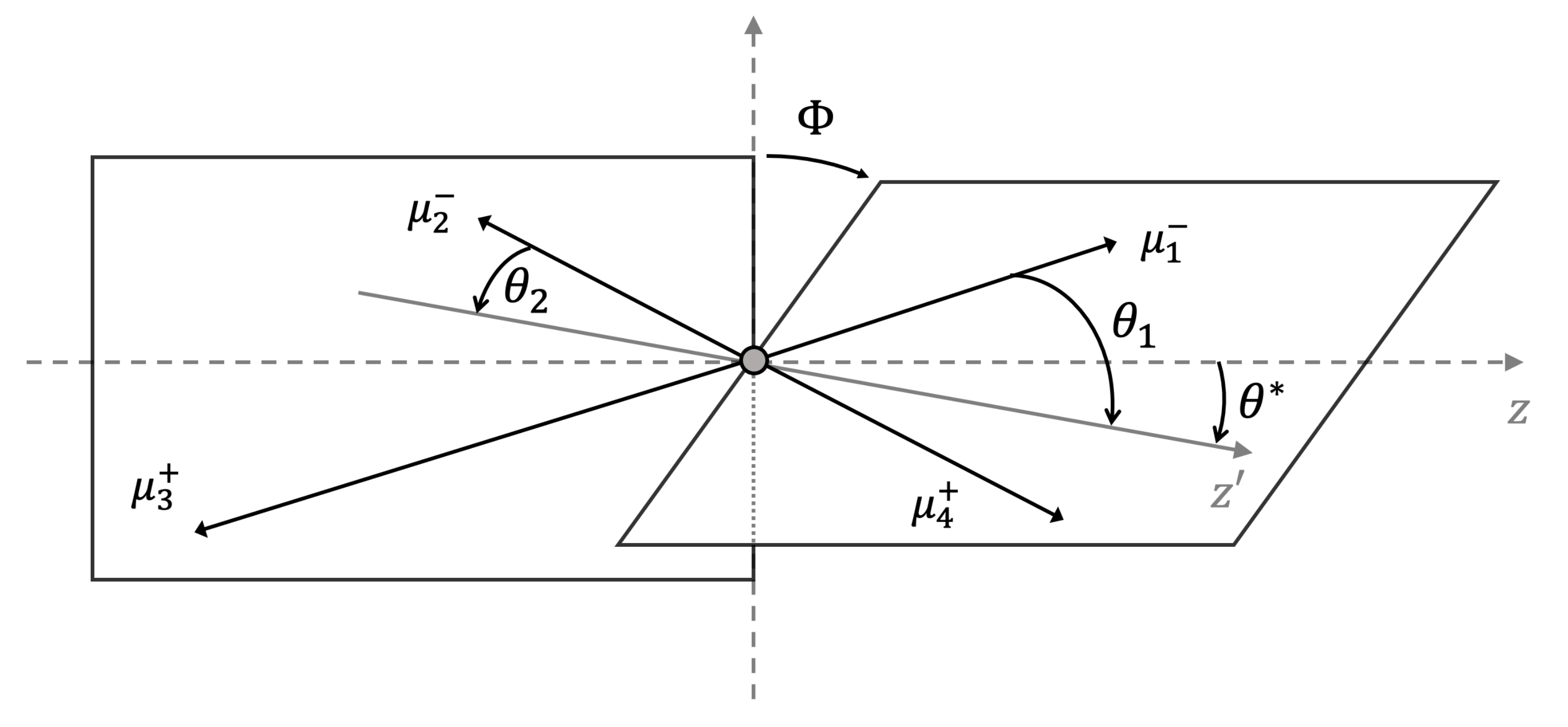}
\caption{Illustration of the angles characterizing the decay $Z\to 4\mu$ in the rest frame of the $Z$ boson. The nomenclature is adapted from~\cite{Gao:2010qx}. $Z'$ denotes the direction of the boost of the highest $p_T$ muon-system. Not shown is the angle $\Phi_1$ which is between the normal vectors of the planes spanned by the $Z$ axis and $Z'$ as well as the one spanned by the highest $p_T$ muons.}
\label{fig:angles}
\end{figure}

The fits in the previous section to the inclusive $Z \to 4 \mu$ measurement and
neutrino-trident production probe two independent combinations of the
three four-muon Wilson coefficients.  With differential measurements
of $Z \to 4 \mu$, all three coefficients can be determined. Enough $Z \to 4 \mu$ events will be
available to allow for differential measurements with high-luminosity LHC data.  Four-lepton final
states are defined by five angles~\cite{Gao:2010qx}: $\theta_1$,
$\theta_2$, $\theta^{*}$, $\Phi_1$, and $\Phi$. We illustrate their definitions in
Fig.~\ref{fig:angles}. When defining these angles we have several choices of
pairing each muon with an anti-muon. Labeling the muon momenta as
$p_1$, $p_2$ with $p_T(p_1) > p_T(p_2)$, and the anti-muons as $p_3$,
$p_4$ with $p_T(p_3) > p_T(p_4)$, we find that if we look at
single-differential distributions, the pairing that gives
the most discrimination between SMEFT-induced deviations and the SM is
$p_1$ with $p_4$ and $p_2$ with $p_3$.  
To demonstrate that this is the optimal pairing we perform simple
one-dimensional fits between the SM and the SMEFT with a single Wilson
coefficient turned on. We define the following test statistic:
\begin{equation}
\label{chisq}
\chi^2 \equiv \sum_{i=1}^{\#  \rm{of \; bins}} {\frac{(N_i^{\rm SMEFT} - N_i^{\rm SM} )^2}{(\sigma_i^{\rm SM})^2}} 
\end{equation}
 where the number of bins is set to 10 and $N_i^{\rm SM} (N_i^{\rm SMEFT})$ stands for the number of SM (SMEFT) events in the $i$th bin.
$ \sigma_i^{\rm SM} =\sqrt{N_i^{SM}}$ represents the standard
deviation of the $i$th bin. For each of the four-muon Wilson
coefficients we use this test statistic to probe the sensitivity of each
variable to deviations for each pairing. The results are shown in
Tables~\ref{tab:chi13} and \ref{tab:chi14}, where we have highlighted
the three most discriminating cases for each pairing. We see that the
$p_1-p_4$ pairing is generally more sensitive, and that the most
discriminating variables are $\theta_1$ and $\theta_2$.

\begin{table}[ht]
\centering
\begin{tabular}{c||c|c|c|c|c}
 ($l_1$, $l_3$) &  $\cos\theta^*$   & $\cos\theta_1$ & $\cos\theta_2$   &  $\Phi_1$ & $\Phi$ \\
\hline\hline
$C_{\substack{ll\\2222}}$      & 39.8  & \textcolor{red}{73.3}  & 18.1   & 9.7    &  15.2    \\

$C_{\substack{ee\\2222}}$   & 37.3  & 41.6   & 14.0   & 17.0 &  16.9 \\

$C_{\substack{le\\2222}}$    & 16.0  & \textcolor{red}{51.0}  & 18.7   & 10.2  & \textcolor{red}{76.3}  \\ 
\hline\hline
\end{tabular}
\caption{$\chi^2$ values for the five single-differential distribution. $l_1$ and $l_3$ ($l_2$ and $l_4$) are grouped together in the same decay plane. The three largest $\chi^2$ values are highlighted.}\label{tab:chi13}
\end{table}

\begin{table}[ht]
\centering
\begin{tabular}{c||c|c|c|c|c}
($l_1$, $l_4$) &  $\cos\theta^*$   & $\cos\theta_1$ & $\cos\theta_2$   &  $\Phi_1$ & $\Phi$ \\
\hline\hline
$C_{\substack{ll\\2222}}$      & 24.6  & 77.8     & 61.6    & 24.1   &  65.3    \\

$C_{\overset{22}{2222}}$    &  11.6 & 44.9   & \textcolor{red}{102.1}       & 21.0    & 69.6   \\

$C_{\substack{le\\2222}}$     & 6.6  & \textcolor{red}{375.2}      &   \textcolor{red}{335.5}  & 25.5 & 48.7 \\ 
\hline\hline
\end{tabular}
\caption{$\chi^2$ values for the five single-differential distribution. $l_1$ and $l_4$ ($l_2$ and $l_3$) are grouped together in the same decay plane here. The three largest $\chi^2$ values are highlighted.
\label{tab:chi14}}
\end{table}
To determine the sensitivity of the future LHC data to four-muon
Wilson coefficients, and its complementarity with neutrino-trident
production, we perform fits to both data sets.  We study both 300 fb$^{-1}$ and 3000 fb$^{-1}$ to mimic future LHC data sets.
We construct a two-dimensional differential distribution based on the variables $\theta_1$ and $\theta_2$, which were found above to be
the most sensitive to SMEFT-induced deviations.  While a more sophisticated multi-variate analysis could potentially improve the results found here, we believe that our approach captures the essence of what can be learned from differential measurements. We define a $\chi^2$ function
as follows:
\begin{equation}
\label{chisq-joint}
\chi^2 \equiv \sum_{i=\# \; \rm of\; bins} {\frac{(N_i^{\rm theo} - N_i^{\rm exp} )^2}{(\sigma_i^{\rm exp})^2}}  + \sum_{j} {\frac{(f_j^{\rm theo} - f_j^{\rm exp} )^2} {(\sigma_j^{\rm exp})^2}} 
\end{equation}
where the first term accounts for predicted future LHC data for $Z\to 4\mu$.  $i$ ranges from 1 to the
number of bins of a given differential distribution. In constructing
our binning we impose the requirement $N_i > 10$ so that we can assume
Gaussian errors. The cuts used in
Ref.~\cite{Sirunyan:2017zjc} are applied. We conservatively use the systematic uncertainty from Ref.~\cite{Sirunyan:2017zjc}, neglecting possible improvements with future LHC data, and assume that it is constant and uncorrelated for all bins. We stress that this is only a simple estimate of the LHC potential, and is meant to motivate more detailed future experimental studies. The statistical uncertainty of the $i$th bin is assumed to be $\sqrt{N_i}$. The second term in
Eq.~(\ref{chisq-joint}) accounts for the neutrino-trident experimental measurements discussed in Section~\ref{sec:ntrident}.  $f_j^{\rm theo}$ denotes the theoretical prediction for the neutrino-trident cross section given in Eq.~(\ref{eq:ntrident}), while the $f_j^{\rm exp}$ are the experimental measurements from CCFR and CHARM-II.  The $\sigma_j^{\rm exp}$ in the denominator denote the experimental errors.
 
\begin{figure}[htbp]
\centering
\includegraphics[width=.6\textwidth]{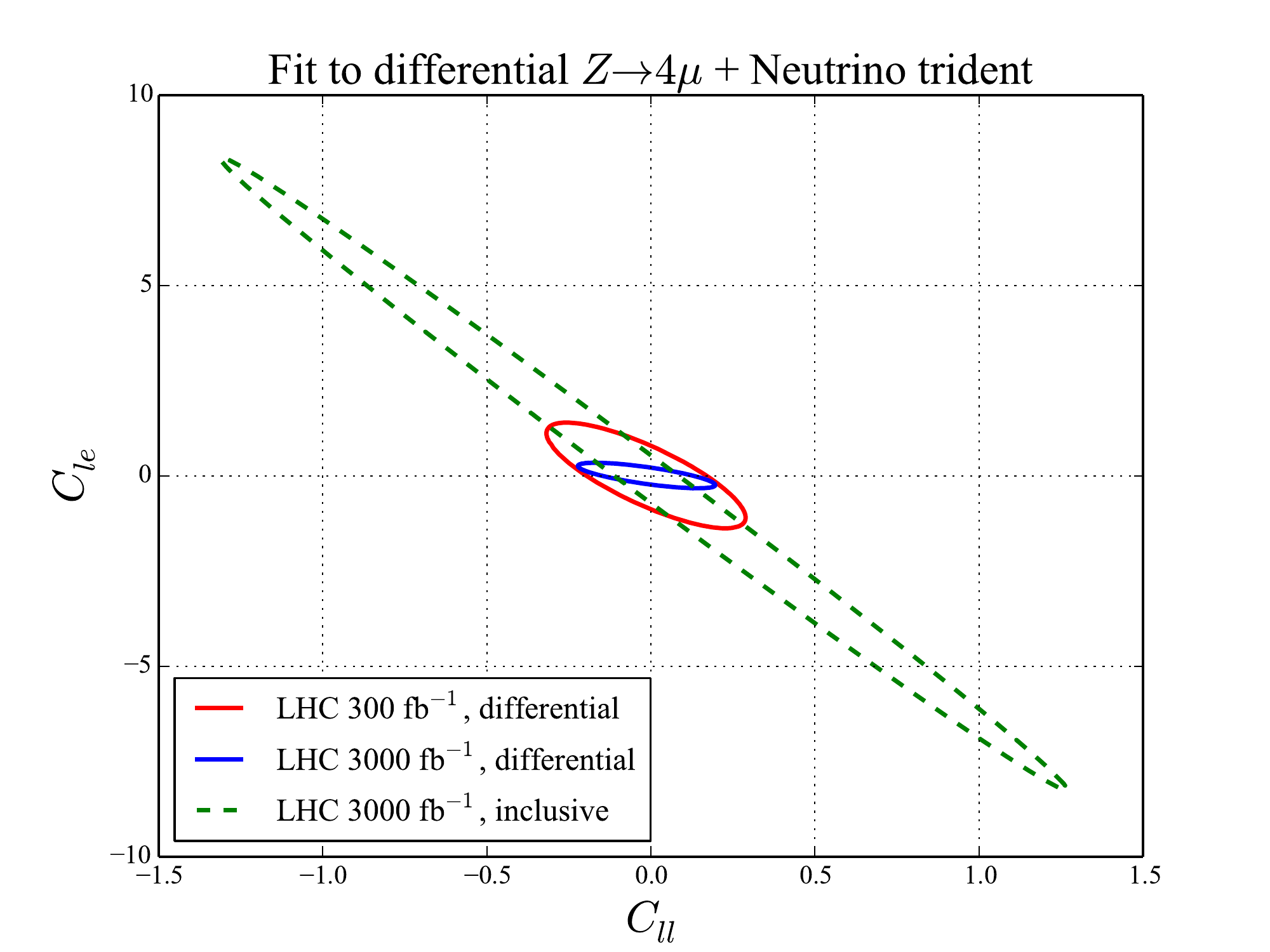}
\caption{68\% C.L. bounds on the combination of differential LHC $Z \to 4 \mu$ data
  and neutrino trident production assuming non-zero $C_{\overset{ll}{2222}}$ and $C_{\overset{le}{2222}}$ }
\label{fig:diffcllcle2}
\end{figure}

To permit simple two-dimensional representations of our results we allow $C_{\substack{ll\\2222}}$ and $C_{\substack{le\\2222}}$ to be
non-zero.  Only a single combination of these parameters can be
determined from neutrino-trident production, so this example will
study how well differential LHC measurements can help break the remaining  degeneracy between Wilson coefficients that occurs given only the inclusive branching ratio measurement.  For comparison we also fit to the inclusive LHC
measurement.  The
results assuming differential LHC measurements are
shown in Fig.~\ref{fig:diffcllcle2}.  For comparison the result assuming only an inclusive branching ratio measurement with 3000 fb$^{-1}$ is shown as well. The improvement going from inclusive to differential
measurements at the LHC is significant, with bounds on
$C_{\substack{le\\2222}}$ improving from ${\cal O}(10)$ to ${\cal
  O}(1)$. This strong improvement is in large part due to the sign of the SMEFT deviations changing in different regions of $(\theta_1,\theta_2)$ space, which is partially averaged out in the inclusive analysis, while the differential analysis resolves the opposite-sign contributions. The flat direction in the $C_{\substack{le\\2222}}$ versus
$C_{\substack{ll\\2222}}$ plane present with just neutrino trident production leads to the elongated shape of the constraint ellipse in this figure. This is removed by the high-luminosity LHC data.

\section{Conclusions}  \label{sec:conc}

In this paper we have studied what can be learned about four-muon
operators in the SMEFT from rare $Z \to 4\mu$ decays at the LHC.
Measurements of this decay mode constrain linear combinations of
Wilson coefficients not accessible in other processes.  We determined
the constraints imposed on these coefficients from current
measurements of the inclusive branching ratio, and showed their
complementarity with existing constraints from neutrino-trident
production.  Future differential measurements of $Z \to 4\mu$ have the
potential to completely determine the four-muon Wilson coefficients in
SMEFT.  We show that strong bounds on all four-muon interactions can
be obtained assuming 3000 fb$^{-1}$ of integrated luminosity at the
LHC.   

We have focused in this paper on the $Z\to 4\mu$ decay since experimental searches for this channel exist. However, measurements of the decay $Z\to2\mu 2\tau$ would probe SMEFT Wilson coefficients 
such as $C_{\overset{ll}{2332}}$, $C_{\overset{le}{2332}}$, and $C_{\overset{ee}{2332}}$, while $Z\to 4\tau$ would probe $C_{\overset{ll}{3333}}$, $C_{\overset{le}{3333}}$, and $C_{\overset{ee}{3333}}$.  Only $C_{\overset{le}{2332}}$ is weakly constrained by $\tau$-decays.  The remaining coefficients are completely untested.  The suggested rare $Z$-decays would provide the first tests 
of this unknown sector of the SMEFT.  To our knowledge these rare $Z$-decays into $\tau$-leptons have not been considered.  However, searches for Higgs bosons into these final states have been performed at the LHC~\cite{Khachatryan:2015nba,Khachatryan:2017mnf}. We encourage the ATLAS and CMS collaborations to perform these searches with future data.

\section*{Acknowledgments}

We thank W.~Hopkins for helpful discussions. R.~B. is supported by the DOE contract DE-AC02-06CH11357.  C.-Y.~C. is
supported by the NSF grant NSF-1740142.  F.~P. and D.~W. are supported
by the DOE grants DE-FG02-91ER40684 and DE-AC02-06CH11357.

\appendix
\section{Vertex shifts}\label{app:A}
We present here the vertex shift factors.  Our conventions for the $Zff$ vertices are such that
\begin{align}
    V^\mu_\textrm{SMEFT}P_{L/R} = V^\mu_\textrm{SM}P_{L/R}\left\{1+\frac{v^2}{\Lambda^2}\delta g^{L/R}_V\right\}.
\end{align}
The Higgs vev can be expressed in term of input parameters as
\begin{equation}
    v^2=\frac{1}{\sqrt{2}G_{\mu}}.
\end{equation}

\noindent
The shift factors that appear in the dimension-6 corrections to the $Z \to 4\mu$ decay width in the $(G_{\mu},M_W,M_Z)$ input scheme are:
\begin{align}
\delta g_{\gamma l_i} =& \frac{1}{4s_W^2} \left\{(C_{\overset{ll}{1221}} + C_{\overset{ll}{2112}} - C_{\overset{\phi l}{11}}^{(3)} - C_{\overset{\phi l}{22}}^{(3)})s_W^2 - C_{\phi D} c_W^2 -4c_Ws_WC_{\phi WB}\right\},\nonumber\\
\delta g_{Z l_i}^L =& \frac{1}{4(1-2s_W^2)}\left\{(C_{\overset{ll}{1221}} + C_{\overset{ll}{2112}} - 2C_{\overset{\phi l}{22}}^{(3)})(1-2s_W^2)+(1+2c_W^2)C_{\phi D} +8c_Ws_WC_{\phi WB}\right.\nonumber\\
&\left.+4C_{\overset{\phi l}{ii}}^{(1)} +2(1+2s_W^2)C_{\overset{\phi l}{ii}}^{(3)}\right\},\nonumber \\
\delta g_{Zl_i}^R =& \frac{1}{4s_W^2} \left\{(C_{\overset{ll}{1221}} + C_{\overset{ll}{2112}} - 2C_{\overset{\phi l}{11}}^{(3)} - 2C_{\overset{\phi l}{22}}^{(3)})s_W^2-(1+c_W^2)C_{\phi D} - 4c_Ws_WC_{\phi WB} -2C_{\overset{\phi e}{ii}}\right\},\nonumber\\
\delta g_{Z\nu_i}^L =& \frac{1}{4}\left\{C_{\overset{ll}{1221}} + C_{\overset{ll}{2112}} - 2 C_{\overset{\phi l}{11}}^{(3)} - 2 C_{\overset{\phi l}{22}}^{(3)} - C_{\phi D} + 4 C_{\overset{\phi l}{ii}}^{(1)} - 4 C_{\overset{\phi l}{ii}}^{(3)}\right\},\nonumber\\
\delta g_{Zu_i}^L = &\frac{1}{4 (4 s_W^2-3)}\left\{(1 - 4 c_W^2) (C_{\overset{ll}{1221}} + C_{\overset{ll}{2112}} - 2 (C_{\overset{\phi l}{11}}^{(3)} + C_{\overset{\phi l}{22}}^{(3)}) - (1 + 
    4 c_W^2) C_{\phi D}\right. \nonumber\\
& \left.- 16 c_W s_W C_{\phi WB} + 12 C_{\overset{\phi q}{ii}}^{(1)} - 12 C_{\substack{\phi q\\ ii}}^{(3)}\right\},\nonumber\\
\delta g_{Zu_i}^R = &\frac{1}{4 s_W^2}\left\{s_W^2 (C_{\overset{ll}{1221}} + C_{\overset{ll}{2112}} - 2 (C_{\overset{\phi l}{11}}^{(3)} + C_{\overset{\phi l}{22}}^{(3)})) - (1 + c_W^2) C_{\phi D} - 
 4 c_W s_W C_{\phi WB} + 3 C_{\overset{\phi u}{ii}}\right\},\nonumber\\
\delta g_{Zd_i}^L = &\frac{1}{4 (1 + 2 c_W^2) s_W}\left\{(1 + 2 c_W^2) (C_{\overset{ll}{1221}} + C_{\overset{ll}{2112}} - 2 (C_{\overset{\phi l}{11}}^{(3)} + C_{\substack{\phi q\\ 22}}^{(3)})) s_W \right.\nonumber\\
&\left.+ 
 8 c_W s_W^2 C_{\phi WB} + (2 c_W^2 -1) s_W C_{\phi D} + 
 12 (C_{\overset{\phi q}{ii}}^{(1)} + C_{\overset{\phi q}{ii}}^{(3)}) s_W\right\},\nonumber\\
\delta g_{Zd_i}^R = &\frac{1}{4 s_W^2}\left\{s_W^2(C_{\overset{ll}{1221}} + C_{\overset{ll}{2112}} - 2 (C_{\overset{\phi l}{11}}^{(3)} + C_{\overset{\phi l}{22}}^{(3)}))  - (1 + c_W^2) C_{\phi D} - 
 4 c_W s_W C_{\phi WB} - 6 C_{\overset{\phi d}{ii}}\right\},\nonumber\\
 \delta g_{Wl_i}^L = &\frac{1}{4}\left\{C_{\overset{ll}{1221}} + C_{\overset{ll}{2112}} - 2 C_{\overset{\phi l}{11}}^{(3)} - 2 C_{\overset{\phi l }{22}}^{(3)} + 4 C_{\overset{\phi l}{ii}}^{(3)}\right\}.
\label{eq:shiftdef}
\end{align}
We have made use of the on-shell definition of the weak mixing angle
\begin{align}
    s_W^2 = 1 - \frac{M_W^2}{M_Z^2} = 1 - c_W^2 .
\end{align}

\section{$Z$ Decay Widths at Leading Order}\label{app:B}
We summarize here all  leading-order SMEFT contributions to the $Z$-decay width that have been used in this paper. Our conventions are such that
\begin{align}
    \Gamma^\textrm{SMEFT}(Z\to f\overline{f}) = \Gamma^\textrm{SM}_{f\overline{f}} + \frac{M_Z^2}{\Lambda^2} \delta \Gamma_{f\overline{f}}
\end{align}
with the following analytic expressions for the SMEFT contributions: 
\begin{align}
    \delta\Gamma_{\nu_i \nu_i} =& \frac{M_Z}{12 \pi}\delta g^L_{Z\nu_i},\nonumber\\
\delta\Gamma_{l_i l_i} =&\frac{M_Z}{12 \pi}\left\{(1 - 2 c_W^2)^2 \delta g_{Zl_i}^L + 4s_W^4 \delta g_{Zl_i}^R\right\},\nonumber\\
\delta\Gamma_{u_i u_i} =&\frac{N_c M_Z}{108 \pi}\left\{16 s_W^4  \delta g_{Zu_i}^R+ (3 - 4 s_W^2)^2 \delta g_{Zu_i}^L \right\},\nonumber\\
\delta\Gamma_{bb} =&\frac{N_c M_Z}{108 \pi}\left\{2 s_W^2 (2 s_W^2 - \beta_b^2 (9 - 4 s_W^2))\delta g_{Zb}^R + (3 - 2 s_W^2) (3 - 2 s_W^2\right.\nonumber\\
&\left.- \beta_b^2 (3 + 4 s_W^2)) \delta g_{Zb}^L-9 \sqrt{2} \beta_b (3 - 4 s_W^2) (s_W C_{\overset{dB}{33}} + c_W C_{\overset{dW}{33}})\right\}\sqrt{1 - 4 \beta_b^2}.
\end{align}
We have assumed a non-vanishing bottom mass, appearing as
\begin{align}
    \beta_b = \frac{m_b}{M_Z}.
\end{align}
The remaining two down-type partial widths can be found by setting $\beta_b = 0$ and appropriately changing the generational indices of the $\{3,3\}$ Wilson coefficients. We give below the numerical expressions for the SMEFT dependence of the decay width for the process $Z\to 4\mu$, obtained with {\tt MadGraph}:
\begin{align}
\Gamma^\textrm{LO}(Z \to 4\mu) = \Gamma^\textrm{LO}_\textrm{SM}(Z \to 4\mu) +\frac{v^2}{\Lambda^2}\bigg\{&0.0260C_{\overset{le}{2222}} + 0.0711C_{\overset{ll}{2222}}+ 0.0444C_{\overset{ee}{2222}} \nonumber\\
& + 15.0C_{\overset{ll}{1221}}+24.0 C_{\phi WB}+6.08C_{\phi D}\nonumber\\
&-0.0062C_{\overset{\phi l}{11}}^{(1)}+12.0C_{\overset{\phi l}{22}}^{(1)}-2.14C_{\overset{\phi l}{11}}^{(3)}\nonumber\\
&+0.445C_{\overset{\phi l}{22}}^{(3)}+0.733C_{\overset{\phi e}{22}}\bigg\}\times 10^{-6}\textrm{GeV}.
\end{align} 
For completeness we also give the result for the process $Z\to 2\mu2e$ below:
\begin{align}
\Gamma^\textrm{LO}(Z \to 2\mu2e) = \Gamma^\textrm{LO}_\textrm{SM}(Z \to 2\mu2e) +\frac{v^2}{\Lambda^2}\bigg\{&29.9 C_{\overset{ll}{1221}} + 0.0752C_{\overset{ll}{1212}}+ 0.180C_{\overset{ee}{1122}} \nonumber\\
& + 0.0840 C_{\overset{ee}{1212}}+ 0.108C_{\overset{le}{1122}} + 0.0769 C_{\overset{le}{1221}}\nonumber\\
&+ 0.0740C_{\substack{le \\ 1212}} + 0.108 C_{\overset{le}{2211}}+46.9C_{\phi WB}\nonumber\\
&+12.1C_{\phi D}+12.1C_{\overset{\phi l}{11}}^{(1)}+12.1C_{\overset{\phi l}{22}}^{(1)}\nonumber\\
&-1.50C_{\overset{\phi l}{11}}^{(3)}-1.50C_{\overset{\phi l}{22}}^{(3)}+0.927C_{\substack{\phi e\\11}}\nonumber\\
&+0.921C_{\substack{\phi e\\22}}\bigg\}\times 10^{-6}\textrm{GeV}.
\end{align} 
The leading-order SM results are
\begin{align}
\Gamma^\textrm{LO}_\textrm{SM}(Z \to 4\mu) = 2.86\times 10^{-6}\,\textrm{GeV} \;\;\;\;\textrm{and}\;\;\;\; \Gamma^\textrm{LO}_\textrm{SM}(Z \to 2\mu2e) = 5.62\times 10^{-6}\,\textrm{GeV}.
\end{align}
%


\end{document}